\documentclass[prl, aps, twocolumn, amsmath, amssymb, superscriptaddress]{revtex4-1}
\usepackage{graphicx,verbatim}
\usepackage{braket}
\usepackage{color}

\usepackage{soul}
\usepackage[normalem]{ulem}
\usepackage[x11names]{xcolor}

\begin{document}

\title{Quantum Monte Carlo in the steady-state}

\author{A.\ Erpenbeck}
\affiliation{
	Department of Physics, University of Michigan, Ann Arbor, Michigan 48109, USA
	}

\author{E.\ Gull}
	\affiliation{
	Department of Physics, University of Michigan, Ann Arbor, Michigan 48109, USA
	}

\author{G.\ Cohen}
	\affiliation{
	The Raymond and Beverley Sackler Center for Computational Molecular and Materials Science, Tel Aviv University, Tel Aviv 6997801, Israel
	}
	\affiliation{
	School of Chemistry, Tel Aviv University, Tel Aviv 6997801, Israel
	}

\date{\today}

\begin{abstract}
    We present a numerically exact steady-state inchworm Monte Carlo method for nonequilibrium quantum impurity models.
    Rather than propagating an initial state to long times, the method is directly formulated in the steady-state.
    This eliminates any need to traverse the transient dynamics and grants access to a much larger range of parameter regimes at vastly reduced computational costs.
    We benchmark the method on equilibrium Green's functions of quantum dots in the noninteracting limit and in the unitary limit of the Kondo regime.
    We then consider correlated materials described with dynamical mean field theory and driven away from equilibrium by a bias voltage.
    We show that the response of a correlated material to a bias voltage differs qualitatively from the splitting of the Kondo resonance observed in bias-driven quantum dots.
\end{abstract}

\maketitle

\paragraph{Introduction.}

    Nonequilibrium driving can have profound effects on strongly correlated materials.
    Experimental probes have revealed intricate phenomena ranging from electrically induced metal--insulator transitions \cite{Stefanovich_Electrical_2000, Kim_Mechanism_2004, Gu_non_Electrical_2013, Stoliar_Universal_2013, Brockman_Subnanosecond_2014, delValle_Subthreshold_2019, del_Valle_Spatiotemporal_2021, delValle_Dynamics_2021}
    to metastable nonequilibrium states \cite{Ogasawara_Ultrafast_2000, Stojchevska_Ultrafast_2014, Zhang_Cooperative_2016, Dean_Ultrafast_2016}
    and driven superconductivity \cite{Fausti_Light_2011, Nicoletti_Optically_2014, Mitrano_Possible_2016, Bao_Light_2022}.
    The rich variety of electronic phases and their sensitive dependence on external fields makes the physics of driven correlated electron systems a promising basis for next-generation nanoscale electronics \cite{Takagi_Emergent_2010}.
    The theoretical description of such physics, however, remains a challenging task.
    
    To study a strongly correlated system out of equilibrium, one may expose it to short external stimuli or quenches, after which time propagation can be probed.
    This is the guiding principle for pump-probe approaches \cite{Iwai_Ultrafast_2003, Perfetti_Time_2006, Wall_Quantum_2011, Fausti_Light_2011, Dean_Polaronic_2011, Ichikawa_Transient_2011, Dietze_X-ray_2014, Konstantinova_Nonequilibrium_2018, Ronchi_Early_2019}.
    Another strategy is driving the correlated material into a time-independent nonequilibrium steady-state, e.g.\ by applying an external bias voltage or temperature gradient.
    This is the approach taken in transport experiments \cite{Cronenwett_Low_2002, Limelette_Universality_2003, Kagawa_Transport_2004, Lee_Electrically_2008, Gu_non_Electrical_2013, delValle_Electrically_2017, Gonzalez_Nanoscale_2020}.

    Theoretical methodologies for the accurate description of transport experiments should ideally
    (i) be able to describe all aspects of correlation physics without introducing spurious artifacts, and
    (ii) be able to access the steady-state.
    The development of numerically exact methods that meet these requirements is an active field of research \cite{Hou_Hierarchical_2014, Erpenbeck_Extending_2018, Zhang_Hierarchical_2020, Chen_Universal_2022, Balzer_Multiconfiguration_2015, Schwarz_Nonequilibrium_2018, Lotem_Renormalized_2020}. 
    Successful approaches have so far relied on time propagation from a tractable initial state, over timescales long enough to make the initial state irrelevant \cite{Balzer_Multiconfiguration_2015, Schwarz_Nonequilibrium_2018, Lotem_Renormalized_2020}. 
    This strategy may become challenging when targeting strongly correlated steady-states, where coherence times can be orders of magnitude longer than the intrinsic timescales in the electronic Hamiltonian \cite{Nordlander_How_1999,Gull_Numerically_2011,Cohen_Numerically_2013}.
    Since most numerically exact methods face an exponential scaling of computational cost with simulation time \cite{White_Real_2004, Vidal_Efficient_2004, Daley_Time_2004, Anders_Real_2005, Tanimura_Stochastic_2006, Welack_The_2006, Jin_Exact_2008, Hartle_Decoherence_2013, Hartle_Transport_2015}, numerically exact simulations of steady-states for correlated systems are often prohibitively expensive.

    In this Letter, we present a numerically exact algorithm, applicable to strongly correlated systems, that provides direct access to the nonequilibrium steady-state of quantum impurity models without propagation from an initial state.
    The method is based on inchworm Quantum Monte Carlo (iQMC) \cite{Cohen_Taming_2015} 
    and its cost scales linearly with the coherence time of the system.
    This enables the investigation of steady-states forming at timescales that are orders of magnitude longer than what is accessible by techniques based on direct time propagation.
    
    We use the equilibrium Anderson impurity model to benchmark our method against analytical limits and state-of-the-art numerical methods, showcasing its potential for quantum transport applications.
    We then apply our method to correlated materials within the equilibrium dynamical mean-field theory (DMFT) framework \cite{Georges_Dynamical_1996}, demonstrating that it can be used to obtain real frequency spectra without the need for analytical continuation \cite{Jarrell_Bayesian_1996, Fei_Nevanlinna_2021}.
    Using nonequilibrium DMFT \cite{Freericks_Nonequilibrium_2006, Aoki_Nonequilibrium_2014, Turkowski_Nonequilibrium_2021}, we finally consider a strongly correlated material placed between two metallic leads and subjected to a bias voltage \cite{Arrigoni_Nonequilibrium_2013, Aoki_Nonequilibrium_2014}. 
    This setup bears resemblance to a quantum dot driven by a bias voltage, as considered by Meir et al. \cite{Meir_Low_1993}, where it was shown that the Kondo peak or Abrikosov--Suhl resonance can be split by the voltage \cite{Wingreen_Anderson_1994, Lebanon_Measuring_2001, DeFranceschi_Out_2002, Leturcq_Probing_2005, Shah_Nonequilibrium_2006, Fritsch_Nonequilibrium_2010, Cohen_Greens_2014, Cohen_Greens_2014b,Fang_Nonequilibrium_2018, Krivenko_Dynamics_2019}.
    We investigate the effect of nonequilibrium driving on the quasi-particle peak, which is the lattice counterpart of the Kondo effect in the strongly correlated metallic regime.
    There, we show that an analogous---though, intriguingly, more subtle---splitting occurs.

\paragraph{Inchworm Monte Carlo in the steady-state.}\label{sec:method}

    Consider a quantum impurity model described by the Hamiltonian $H = H_\text{I} + H_\text{B} + H_\text{IB}$, consisting of an interacting impurity $H_\text{I}$, a noninteracting bath $H_\text{B}$, and the coupling or hybridization between them $H_\text{IB}$.
    The central object of our method is the restricted propagator between two times on opposite branches of the Keldysh contour,
    $\Phi_\alpha^\beta(t,t')	=	\text{Tr}_\text{B} \left\lbrace \rho_\text{B}
								\bra{\alpha}e^{iHt}\ket{\beta} \bra{\beta}e^{-iHt'}\ket{\alpha}
								\right\rbrace$.
	Here, $\alpha$ and $\beta$ are states in the impurity subspace, with $\alpha$ taking the role of the impurity's initial condition.
	$\rho_\text{B}$ is the initial bath density matrix  and $\text{Tr}_\text{B}$ denotes a trace over the bath degrees of freedom.
	A detailed discussion of these propagators and their relationship to physical observables is given in Ref.~\onlinecite{Cohen_Greens_2014, Erpenbeck_Resolving_2021}.
	
	We calculate the restricted propagators using an expansion in the impurity-bath hybridization $H_\text{IB}$.
	The resulting high-dimensional integral expression for $\Phi_\alpha^\beta(t,t')$ can be evaluated by means of Monte Carlo methods \cite{Muhlbacher_Real_2008, Werner_Diagrammatic_2009, Schiro_Real_2009, Gull_Continuous_2011}.
	A direct summation of all contributions results in the dynamical sign problem, an exponential growth in the required computational resources with simulation time \cite{Muhlbacher_Real_2008, Werner_Diagrammatic_2009, Schiro_Real_2010, Gull_Bold_2010, Cohen_Greens_2014, Gull_Numerically_2011, Antipov_Voltage_2016} that effectively limits this approach to short times.
	The dynamical sign problem can often be overcome by iQMC \cite{Cohen_Taming_2015, Antipov_Currents_2017, Chen_Inchworm_2017,
	Chen_Inchworm_2017, Chen_Inchworm_2017_2, Cai_Inchworm_2020, Cai_Numerical_2020}.
	Instead of calculating the restricted propagator directly, iQMC makes optimal use of the information incorporated in short-time dynamics to construct more efficient expansions for longer times, substantially reducing the number of terms that need to be evaluated.
	However, as iQMC requires knowledge of restricted propagators at all previous times, the algorithm scales at least quadratically with simulation time.

    For the purpose of this work, the established two-time iQMC scheme can be considered to be a map, $F_{\text{inch}}$, from the  set of known restricted propagators up to the times $t_1$ and $t_2$,
    to the restricted propagator at (slightly) larger times $t$ and $t'$, with
    $t_1\leq t$ and $t_2 \leq t'$.
    This can be written as follows:
    \begin{eqnarray}
        F_{\text{inch}}: \left\{ \left. \Phi_\alpha^\beta(\tau,\tau') \right| \tau\leq t_1, \tau'\leq t_2  \right\} &\rightarrow& \Phi_\alpha^\beta(t,t').
    \end{eqnarray}
    As the map $F_{\text{inch}}$ is expressed in terms of known propagators, only a subclass of the summands that would appear in a bare hybridization expansion need to be included; these contributions are referred to as \emph{inchworm proper} \cite{Cohen_Taming_2015, Chen_Inchworm_2017, Chen_Inchworm_2017_2}.
    Given that $\Phi_\alpha^\beta(0,0)=\delta_{\alpha\beta}$, the iQMC method uses the map $F_{\text{inch}}$ to propagate the restricted propagator forward in time, sequentially increasing each of $t$ and $t'$ in a set of small steps until both are sufficiently large, and resulting in the aforementioned quadratic scaling \cite{Cohen_Taming_2015}.
    
    Notably, once the steady-state is reached the propagator becomes characterized by two simplifying conditions.
    It is (i) independent of the impurity's initial condition $\alpha$, and (ii) invariant to propagation in the direction $t,t' \rightarrow t+\Delta t, t'+\Delta t$; or, equivalently, dependent only on the time difference $t-t'$ rather than on the explicit values of $t$ and $t'$.
    We use these facts to formulate an iQMC method directly in the steady-state, by explicitly seeking a solution of the form $\Phi_\alpha^\beta(t,t')=\Phi^\beta(t-t')$.
    From the previous discussion, it follows that a steady-state $\Phi^\beta(t-t')$ must be a fixed point of the iQMC procedure:
    \begin{eqnarray}
        F_{\text{inch}}: \left\{ \Phi^\beta(t-t') \right\} &\rightarrow& \Phi^\beta(t-t'). \label{eq:self-cons}
    \end{eqnarray}
    That is, given the exact propagator at \emph{all} time differences, the iQMC method will generate the same propagator at \emph{each} time difference.
    We use this property as a self-consistency condition.
    The construction of the iQMC map $F_{\text{inch}}$, the associated diagrammatic representation and the notion of inchworm proper diagrams remain identical to the time-dependent iQMC scheme, which is discussed in detail in Refs.~\onlinecite{Cohen_Taming_2015, Antipov_Currents_2017, Chen_Inchworm_2017}.
    Note that the self-consistency condition we described exists within iQMC, but not within the bare hybridization expansion \cite{Muhlbacher_Real_2008, Werner_Diagrammatic_2009, Schiro_Real_2009, Gull_Continuous_2011}, because in iQMC the restricted propagator for a given time is expressed in terms of restricted propagators at previous times.

    In practice, we start the self-consistency cycle from some initial guess $\Phi^\beta_0(t-t')$ (we used a low-order perturbative result) and iterate Eq.~(\ref{eq:self-cons}) until the last two results are indistinguishable to within some tolerance.
    Typically, we converge within $100–200$ iterations.
    Since numerically we cannot work with all possible values of $t-t'$, only time intervals $\left| t-t' \right| \leq t_\text{cutoff}$ with some cutoff time $t_\text{cutoff}$ are used; the latter is then a numerical parameter that must be converged.
    The steady-state propagator depends only on the time difference and a single many-body state index in the impurity subspace, and therefore our approach scales linearly with the coherence time and with the dimension of the impurity Hilbert space (which grows exponentially with the number of impurity orbitals).

    Once $\Phi^\beta(t-t')$ is known, the steady-state Green's functions (GFs) are obtained from existing techniques \cite{Cohen_Greens_2014,Antipov_Currents_2017}.
    In particular, we calculate the retarded steady-state GF, $G_{\sigma\sigma'}^r(\Delta t)=-i\theta(\Delta t) \braket{d_{\sigma}(\Delta t)d_{\sigma'}^\dagger(0)+d_{\sigma'}^\dagger(0)d_{\sigma}(\Delta t)}$.
    The spectral function is then given by the Fourier transform of the retarded GF, $A_{\sigma\sigma'}(\epsilon) = -\text{Im}\lbrace G_{\sigma\sigma'}^r(\epsilon) \rbrace$.

\paragraph{Anderson impurity model.}\label{sec:model}
    
    We present results for the Anderson impurity model,
        $H_\text{I} = 	  \sum_{\sigma} \epsilon_0 d_\sigma^\dagger d_\sigma + U d_\uparrow^\dagger d_\uparrow d_\downarrow^\dagger d_\downarrow$,
        $H_\text{B} =      \sum_{\sigma k} \epsilon_{k} a_{k\sigma}^\dagger a_{k\sigma}$ 
        and 
        $H_\text{IB} =     \sum_l\sum_{k\in l}\sum_\sigma\left( V_{k} a_{k\sigma}^\dagger d_\sigma + \text{h.c.} \right)$. 
    Here, $\epsilon_0$ is the on-site energy on the impurity and $U$ is the Coulomb interaction strength.
    $d_\sigma$ and $d^\dagger_\sigma$, respectively, are annihilation and creation operators with spin $\sigma\in\lbrace \uparrow, \downarrow  \rbrace$ on the dot. $a_{k\sigma}$ and $a^\dagger_{k\sigma}$ are their counterparts on the bath orbitals with energy $\epsilon_k$.
    The dot--bath coupling for bath $\ell$ is characterized by the coupling strength function $\Gamma_{\ell}(\epsilon) = 2\pi \sum_{k\in \ell} |V_{k}|^2 \delta(\epsilon-\epsilon_{k})$.
    $\Gamma_{\ell}(\epsilon)$ is used to model either leads or an effective DMFT bath, each of which is initially at equilibrium \cite{Aoki_Nonequilibrium_2014}.
    Applying different chemical potentials $\mu_\ell$ to different leads $\ell$ generates nonequilibrium steady-states.
    Since all the observables to be considered here are spin-independent, spin indices will be suppressed for the remainder of this work.

\paragraph{Benchmark I: resonant level model.}\label{sec:RLM}
    \begin{figure}
	\includegraphics{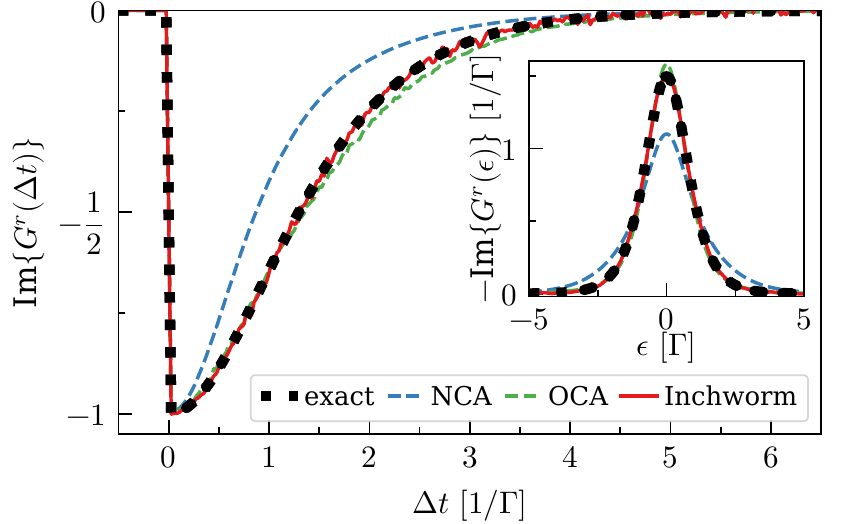}
	\caption{
		Resonant level model: steady-state retarded GF in real time and spectral function in real frequency (inset).
		Parameters are $\epsilon_0 = U = 0$ at temperature $T=\Gamma$, with the impurity coupled to a single, flat bath with bandwidth $\omega_c=1.5\Gamma$ and cutoff width $1/\eta=\Gamma$: $\Gamma(\epsilon) = \Gamma/[(1+e^{\eta((\epsilon-\mu)-\omega_c)})(1+e^{-\eta((\epsilon-\mu)+\omega_c)})]$ and $\mu=0$.
		The iQMC scheme converges at order 4(6) for the propagator (GF).}
        \label{fig:noninteracting}
	\end{figure}
    We consider the noninteracting limit $U=\epsilon_0=0$ at equilibrium coupled to a single bath. While this system is exactly solvable \cite{Bruus_Many_2004, Haug_Qauntum_2008}, it is a challenging benchmark for hybridization expansions.
    Fig.~\ref{fig:noninteracting} shows the retarded GF and the spectral function obtained from steady-state iQMC at different cutoff orders, together with the analytic result (dotted black curve).
    The noncrossing approximation (NCA) and one-crossing approximation (OCA) represent the lowest and next-to-lowest order truncation of the inchworm hybridization expansion \cite{Bickers_Review_1987, Pruschke_Anderson_1989, Pruschke_Hubbard_1993, Haule_Anderson_2001, Eckstein_Nonequilibrium_2010, Cohen_Greens_2014, Erpenbeck_Revealing_2021}.
    When beyond-OCA diagrams are added, the propagator (not shown) stabilizes at fourth order, in the sense that including higher orders does not alter the results within the Monte Carlo error.
    In the same sense, the GFs stabilize at sixth order.
    The data is consistent with the exact result to within the Monte Carlo error.
    A time-dependent iQMC calculation \cite{Cohen_Taming_2015, Boag_Inclusion_2018} performed on the time-grid used for the steady-state simulation of Fig.~\ref{fig:noninteracting} would have a computational cost approximately $\sim$1000 times larger than the steady-state formulation, not including the time needed to overcome the transient dynamics from a known initial state.

\paragraph{Benchmarks II: Kondo regime.}\label{sec:Kondo}
    \begin{figure}
        \includegraphics{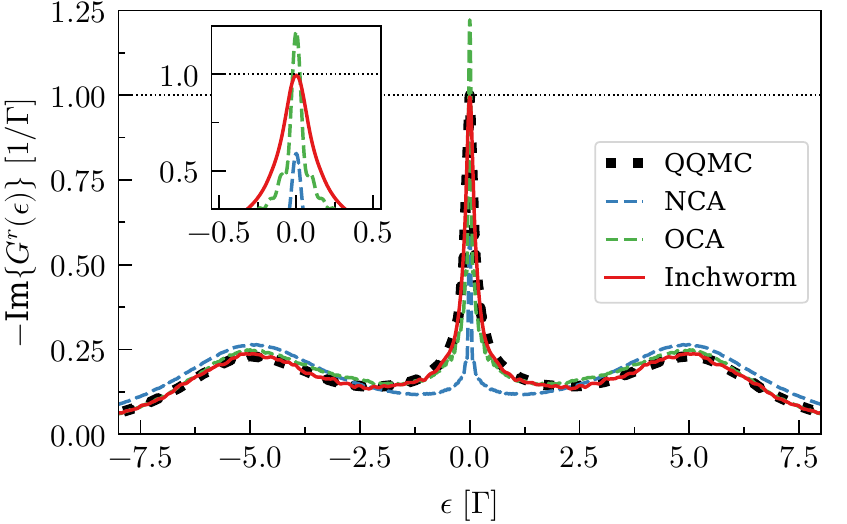}
        \caption{
            Steady-state spectral function in the Kondo regime.
            The bath consists of a single semi-elliptic band $\Gamma(\epsilon) =  \frac{\Gamma}{\Delta} \sqrt{\Delta^2 - \epsilon^2}$ at $\mu=0$, where $|\epsilon|<\Delta$, with $\Delta=11.476\Gamma$.
            Parameters are particle-hole symmetric with $U = -2\epsilon_0 = 8\Gamma$, the temperature is $T=10^{-8}\Gamma$.
            The iQMC scheme converges at order 15(9) for the propagator (GF).
            QQMC data is extracted from Ref.~\cite{Bertrand_Quantum_2021}.
            Inset: low energies and Kondo resonance.
        }
        \label{fig:Kondo}
    \end{figure}
    Providing an accurate description of Kondo physics and capturing the associated Abrikosov--Suhl resonance in the spectral function is a paradigmatic benchmark for methods applicable to correlated systems. 
    In Fig.~\ref{fig:Kondo}, we consider parameters that were recently studied using two state-of-the-art algorithms: the quantum quasi Monte Carlo (QQMC) \cite{Macek_Quantum_2020,Bertrand_Quantum_2021} and the fork tensor product state (FTPS) \cite{Bauernfeind_Fork_2017} methods.
    We compare the spectral function of the Anderson impurity model in the Kondo regime, as calculated by steady-state iQMC, to the QQMC data provided in Ref.~\cite{Bertrand_Quantum_2021} and validated there against FTPS.
    The methods agree within the Monte Carlo error.
    The spectral function exhibits a sharp Kondo peak at the Fermi level and two Hubbard bands centered around $\epsilon\sim\pm U/2$. 
    Within Monte Carlo errors, the iQMC data reproduces the Friedel sum rule  $\text{Im}[G^r(\epsilon=0)]=-1/\Gamma$ \cite{Hewson_Kondo_1997} and resolves the low energy frequency dependence well (see inset).
    In contrast, while the NCA and OCA capture the main features, they fail to provide quantitative results, especially at low energies.
    In this example, the NCA underestimates the height of the Kondo resonance, whereas the OCA overestimates it.
    We note that obtaining converged NCA and OCA results is a nontrivial task, as these low order approximations may overestimate the coherence time by an order of magnitude \cite{Gull_Numerically_2011}, making them prone to finite-time artifacts such as oscillations in the spectral function.

\paragraph{Equilibrium DMFT.}\label{sec:DMFT}
    Interacting lattice models can be mapped self-consistently onto effective impurity models with DMFT \cite{Georges_Dynamical_1996, Vollhardt_Dynamical_2012, Vollhardt_Dynamical_2012_2, Aoki_Nonequilibrium_2014}. 
    The method is in general approximate, but becomes exact for the infinite coordination number Bethe lattice \cite{Georges_Dynamical_1996}.
    In order to contrast the spectral features of an impurity with that of a correlated material, we consider three different scenarios, all with either two leads at equal chemical potentials, $\mu_\text{L}=\mu_\text{R}=0$; or no leads.
    The scenarios are:
    (i) an impurity coupled to noninteracting leads, as in the previous paragraphs;
    (ii) an isolated Bethe lattice in the infinite coordination number limit, describing a correlated material in equilibrium \cite{Metzner_Correlated_1989, Muller_Correlated_1989, Georges_Hubbard_1992}; and
    (iii) an infinite-coordination Bethe lattice coupled to noninteracting leads, describing a correlated material in a junction \cite{Okamoto_Nonlinear_2008, Arrigoni_Nonequilibrium_2013, Aoki_Nonequilibrium_2014,Kleinhenz_Dynamic_2020}.
    \begin{figure}[t]
        \includegraphics{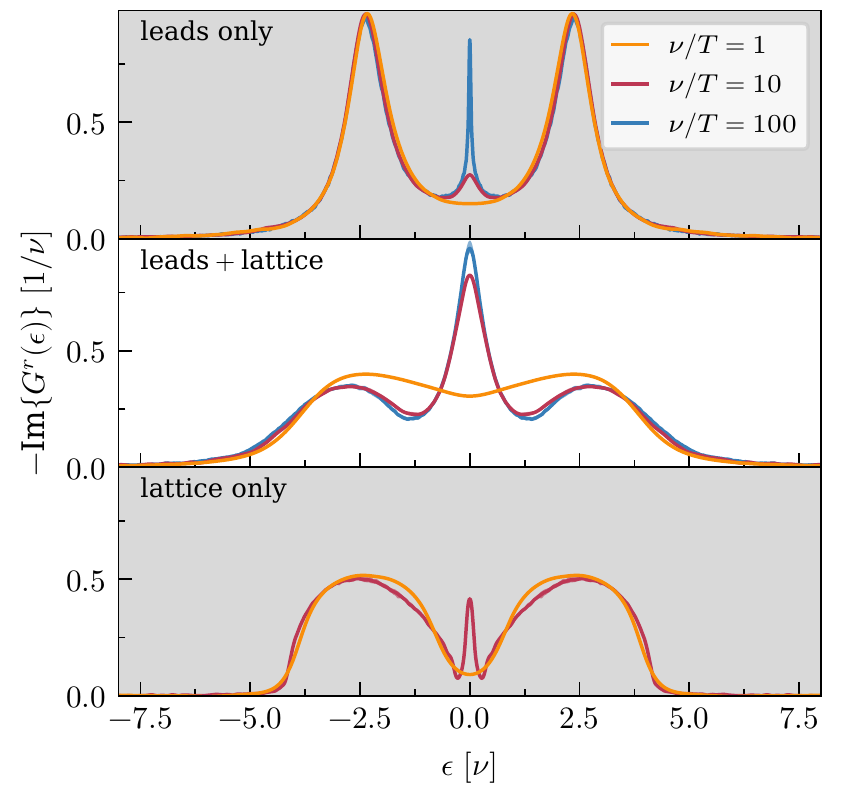}
        \caption{
            Interacting local equilibrium spectral function at different temperatures for an impurity coupled to leads (top), an isolated correlated material (bottom), and a correlated material coupled to leads (middle). 
            The material is modeled as an infinite-dimensional Bethe lattice with inter-site hopping $\nu$, which we use as the overall energy scale; other parameters: $U = -2\epsilon_0 = 4.6\nu$. 
            The leads are parametrized by wide, flat bands with smooth cutoffs: $\Gamma_\ell(\epsilon) = \Gamma/[(1+e^{\eta((\epsilon-\mu_\ell)-\omega_c)})(1+e^{-\eta((\epsilon-\mu_\ell)+\omega_c)})]$, with $\Gamma=0.125\nu$, $\eta=10/\nu$, $\omega_c=10\nu$, $\mu_\ell=0$ and $\ell\in\lbrace\text{L,R}\rbrace$.
            To estimate the error, we plot the average and the standard deviation of three consecutive DMFT iterations (middle/bottom panel), where typical errors are of the size $3\cdot 10^{-2}/\nu$.}
        \label{fig:DMFT_1}
    \end{figure}

    Fig.~\ref{fig:DMFT_1} presents iQMC results for the equilibrium spectral functions for these three scenarios at different temperatures. 
    All three systems exhibit a coherence peak at the Fermi energy at low temperatures, and two Hubbard bands centered around $\epsilon\simeq\pm U/2$.
    In all cases, increasing the temperature results in the melting of the coherence peak and a redistribution of its spectral weight to higher energies.
    In the impurity case (top panel), features are sharpened by a relatively weak coupling to the leads.
    In the isolated lattice (bottom panel), features are more rounded, but the band edge is sharper.
    This case is analogous to standard (usually imaginary time) DMFT calculations, and we have validated our results against numerically exact imaginary-time Monte Carlo data (not shown) \cite{Gull_Continuous_2008}.
    Finally, the lattice coupled to a junction exhibits the most pronounced correlation effect, since the effective impurity couples to both the correlated lattice and the leads, and as a result has a higher Kondo temperature \cite{Hewson_Kondo_1997}.

\paragraph{Nonequilibrium DMFT.}\label{sec:Noneq-DMFT}
    The infinite-coordination Bethe lattice coupled to leads can be driven away from equilibrium by applying a bias voltage $\phi=\mu_\text{L}-\mu_\text{R}$ between the two leads, making them an electronic junction.
    We apply a symmetric bias voltage, $\mu_\text{L}=-\mu_\text{R}$.
    The model remains exactly solvable by way of nonequilibrium DMFT \cite{Freericks_Nonequilibrium_2006, Aoki_Nonequilibrium_2014, Turkowski_Nonequilibrium_2021}.
    Previous studies of similar systems were restricted to approximate impurity solvers \cite{Okamoto_Nonlinear_2008, Arrigoni_Nonequilibrium_2013, Tsuji_Nonequilibrium_2013, Titvinidze_Transport_2015, Kleinhenz_Dynamic_2020}.
    \begin{figure}
        \includegraphics{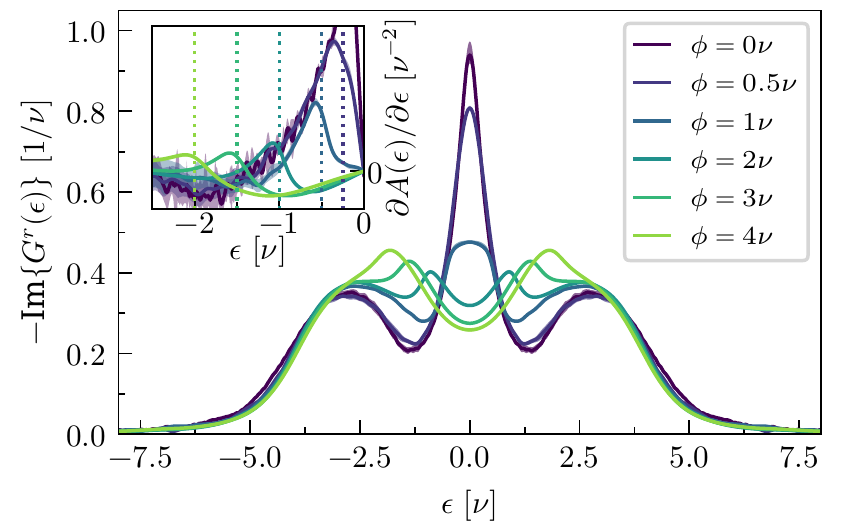}
        \caption{
            Nonequilibrium spectral function for a correlated material in a junction for different bias voltages $\phi$.
            The parameters are as in the middle panel of Fig.~\ref{fig:DMFT_1}, with temperature $T=\nu/100$.
            The average and the standard deviation of three consecutive DMFT iterations are shown to provide an error estimate, where typical errors are of the size $3\cdot 10^{-2}/\nu$.
            Inset: derivative of the spectral function with respect to energy. Vertical dashed lines: value of the chemical potential in the right lead.
            Error estimates are provided as the standard deviation of three consecutive DMFT iterations.
        }
        \label{fig:DMFT_2}
    \end{figure}
    
    Fig.~\ref{fig:DMFT_2} shows the spectral function at different bias voltages $\phi$, with $\phi=0$ corresponding to the equilibrium case in Fig.~\ref{fig:DMFT_1}.
    At low bias $\phi \lesssim 1\nu$, we observe a shrinking of the quasi-particle peak and a formation of a plateau centered around $\epsilon=0$, with its width determined by $\phi$; peaks in the derivative of the spectral function with respect to energy appear at $\epsilon\approx\mu_\text{L/R}$ (see inset in Fig.~\ref{fig:DMFT_2}).
    At higher bias, $\phi \gtrsim 2\nu$, the peak splits into two short peaks centered at $\epsilon=\mu_\text{L/R}$, which are narrower than half the width of the plateau.
    The splitting is as one would expect in quantum dot junctions \cite{Meir_Low_1993,DeFranceschi_Out_2002, Leturcq_Probing_2005, Shah_Nonequilibrium_2006, Han_Imaginary_2007, Anders_Steady_2008, Fritsch_Nonequilibrium_2010, Cohen_Greens_2014, Dorda_Auxiliary_2015, Fugger_Nonequilibrium_2018, Krivenko_Dynamics_2019, Fugger_Nonequilibrium_2020}.
    The formation of the plateau at intermediate values of $\phi$, however, is surprising.
    It indicates that the correlated material stabilizes its equivalent of the Kondo resonance and pins it to the equilibrium chemical potential, a mechanism not present in quantum dot junctions.
    We speculate that the latter effect can be understood from theoretical considerations by investigating the effectively enhanced hybridization weight at low frequencies.
    However, to our knowledge it has not been previously predicted.

\paragraph{Conclusion.}\label{sec:conclusion}
    We presented a Monte Carlo method for quantum impurity models that is formulated directly in the steady-state and is applicable in and out of equilibrium.
    The method, which is based on the inchworm hybridization expansion, can be used to describe transport through quantum dot junctions directly, and strongly correlated materials by way of the DMFT.
    Our formulation utilizes the fact that the two-time structure of restricted propagators on the Keldysh contour can be reduced to a time-difference representation in the steady-state, which enables their evaluation at a computational cost that scales linearly with the coherence time of the system.
    The scheme then provides a self-consistency condition for the restricted propagators, which is solved iteratively.
    We implemented and benchmarked the method for the Anderson impurity model and showed that it can be used to describe a strongly correlated material using DMFT, providing real frequency data without resorting to analytical continuation.
    We then investigated a correlated material in a junction between two metallic leads and driven out of nonequilibrium by a bias voltage.
    There, we found a bias induced splitting of the quasi-particle peak.
    However, this splitting occurs at higher voltages than one might expect from our understanding of transport through quantum dots, and is preceded by the formation of a spectral plateau with enhanced low-energy transport at intermediate voltages.
    Our method promises to enable the numerically exact treatment of a wide variety of problems spanning quantum transport, equilibrium materials science and novel types of strongly correlated nonequilibrium effects that are of current experimental interest.

\paragraph{Acknowledgments.}
    We thank J.\ Kleinhenz and Y.\ Yu for helpful discussions.
    A.E.~was funded by the Deutsche Forschungsgemeinschaft (DFG, German Research Foundation) -- 453644843, and by the Raymond and Beverly Sackler Center for Computational Molecular and Materials Science, Tel Aviv University. 
    E.G.~was supported by the Department of Energy via DE-SC-0022088.
    This material is based upon work supported by the U.S. Department of Energy, Office of Science, Office of Advanced Scientific Computing Research and Office of Basic Energy Sciences, Scientific Discovery through Advanced Computing (SciDAC) program under Award Number DE‐SC0022088.
    This research used resources of the National Energy Research Scientific Computing Center, a DOE Office of Science User Facility supported by the Office of Science of the U.S. Department of Energy
    under Contract No. DE-AC02-05CH11231 using NERSC award BES-ERCAP0021805.
    G.C.~acknowledges support by the Israel Science Foundation (Grants No.~2902/21 and 218/19) and by the PAZY foundation (Grant No.~308/19).

\bibliography{bib.bib}

\end{document}